\documentclass{moriond}

\usepackage{graphicx}
\usepackage{amsmath,amssymb}
\usepackage{xspace}

\newcommand{\TeV}{Te\kern -0.1em V\xspace}
\newcommand{\GeV}{Ge\kern -0.1em V\xspace}
\newcommand{\sqn}{\sqrt{s_{{}_\mathrm{NN}}}\xspace}
\newcommand{\sqs}{\sqrt{s}\xspace}
\newcommand{\rppb}{R_{\mathrm{pPb}}\xspace}
\newcommand{\ppb}{\textit{p}+Pb\xspace}
\newcommand{\pp}{\textit{pp}\xspace}
\newcommand{\Lint}{L_\mathrm{int}}
\newcommand{\pT}{p_\mathrm{T}}
\newcommand{\ET}{E_\mathrm{T}}
\newcommand{\avgTpb}{\langle T_\mathrm{Pb} \rangle}
\newcommand{\Tpb}{T_\mathrm{Pb}}
\newcommand{\dz}{d_0}
\newcommand{\zz}{z_0 \sin \theta}
\newcommand{\ystar}{y^{\star}}

\newcommand{\Photo}{}

\begin{document}

\title{Measurements of charged particle spectra and nuclear modification factor in p+Pb collisions with the ATLAS detector}

\author{Petr Balek (for the ATLAS Collaboration)}
\address{IPNP, Faculty of Mathematics and Physics, Charles University in Prague, V Hole\v{s}ovi\v{c}k\'{a}ch 2,  180 00 Prague,  Czech Republic}

\maketitle

\begin{abstract}

The ATLAS detector at the LHC obtained the sample of \ppb data at $\sqn=5.02$\,\TeV with integrated luminosity of 25\,nb${}^{-1}$, which can be compared to the \pp data obtained by interpolating \pp measurements at $\sqs=2.76$\,\TeV and 7\,\TeV. Due to the excellent capabilities of the ATLAS detector, and its stable operation in heavy ion as well as proton--proton physics runs, the data allow measurements of the nuclear modification factor, ratio of heavy ion charged particle spectra divided by \pp reference, in different centrality classes over a wide range of rapidity. The charged particle nuclear modification factor is found to vary significantly as a function of transverse momentum with a stronger dependence in more peripheral collisions.

\hspace*{1cm}

\noindent \textit{keywords}: proton--lead collision, nuclear modification factor
\end{abstract}

\section{Introduction}

The partonic structure of nuclei has been studied extensively in deep inelastic scattering experiments \cite{Arneodo:1992wf}. These measurements have established differences in the parton distribution functions (PDFs) in nuclei from those expected from an incoherent superposition of nucleons. These differences include the phenomena of shadowing, anti-shadowing \cite{Salgado:2011wc} and the EMC effect \cite{Aubert1983275}. These observations have inspired significant theoretical work (e.g.~\cite{Albacete:2012xq,Barnafoldi:2011px,Kharzeev:2003wz}), but to date the underlying physics of these phenomena is not completely understood. Proton--ion collisions at the LHC are capable of probing the nucleus at scales that have not been experimentally accessible before. This includes particle production from hard scattering processes with large transverse momentum transfer, $q^2$, which may provide crucial insight into the large-$q^2$ structure of the nucleus or even reveal previously unobserved phenomena.

\section{Analysis}

Nuclear modification effects are studied by comparing observables in heavy ion collisions with the corresponding observables obtained in \pp collisions. The rates of hard scattering processes in heavy ion collisions are enhanced relative to \pp collisions due to the increased flux of partons per collision, i.e. the nucleus contains many nucleons capable of participating in a hard scattering. This enhancement depends on the thickness of the target nucleus, $\Tpb$, and on the impact parameter of the projectile. It can be inferred from geometric models such as the Glauber model \cite{Alver:2008aq}. The particle yields in \ppb collisions scaled by the average $\Tpb$ can then be compared directly to the cross section for the same process obtained in \pp collisions:

\begin{equation}
\rppb(\pT) = \frac{1}{\avgTpb}\frac{1/N_\mathrm{evt}~\mathrm{d}^2 N_\mathrm{\footnotesize p+Pb} / \mathrm{d}y \mathrm{d}\pT}{\mathrm{d}^2\mathrm{\sigma}_\mathrm{\footnotesize pp} / \mathrm{d}y \mathrm{d}\pT},
  \label{eq:rpA}
\end{equation}

\noindent where $\rppb$ is called the nuclear modification factor, $1/N_\mathrm{evt}~\mathrm{d}^2 N_\mathrm{\footnotesize p+Pb} / \mathrm{d}y \mathrm{d}\pT$ is the per-event charge particle yield in \ppb collisions measured differentially in transverse momentum ($\pT$) and rapidity ($y$), and $\mathrm{d}^2\mathrm{\sigma}_\mathrm{\footnotesize pp} / \mathrm{d}y \mathrm{d}\pT$ is the corresponding differential cross section measured in \pp collisions.

This analysis \cite{ATLAS-CONF-2014-029} uses \ppb data collected with the ATLAS detector \cite{atlasref} in 2013 at $\sqn=5.02$\,\TeV with a longitudinal rapidity boost of 0.465 relative to the ATLAS laboratory frame. Further, the analysis uses \pp data with $\sqs=2.76$\,\TeV recorded in 2013 and data with $\sqs=7$\,\TeV recorded in 2010 and 2011. The \ppb data have integrated luminosity ($\Lint$) of 25\,nb${}^{-1}$, the \pp data with $\sqs=2.76$\,\TeV and 7\,\TeV have $\Lint \approx 4.0$\,pb${}^{-1}$ and 156\,pb${}^{-1}$, respectively. 

The events used in this analysis are selected from several triggered samples. A minimum bias (MB) samples were obtained requiring signal in the Minimum Bias Trigger Scintilator. For jet triggered samples, jets were reconstructed in events that passed the MB requirement using the anti-k${}_\mathrm{T}$ algorithm \cite{Cacciari:2008gp} with the distance parameter $R=0.4$. Events were selected by the jet trigger if they contained jets with $\ET$ above a certain threshold. Multiple thresholds were defined. Only a fraction of all events which fired a trigger was randomly selected to be recorded for further analysis. This fraction was set differently for each trigger. Total spectra were obtained by merging spectra from MB and jet triggers, all of them scaled by reciprocal of the fraction to obtain MB spectra up to high $\pT$.

The \ppb event centrality classes are defined in terms of percentiles of the total transverse energy measured by the forward calorimeters in the pseudorapidity range $3.1<\eta<4.9$ in the lead-going direction. The $\avgTpb$ values are estimated from the Glauber model \cite{glauber} for each of the studied centrality class.

To study the effects of the detector response on the measurement, Monte Carlo (MC) simulation samples were produced using the PYTHIA event generator \cite{Sjostrand:2006za}. MC samples correspond to \ppb and \pp collisions of the same energy, the same detector configuration and the same beam direction. The MC samples were processed with the same algorithm as the data. For \ppb, a sample of \pp collisions at $\sqs = 5.02$\,\TeV was simulated, but with the output boosted to match the boost due to the asymmetry of the collisions. This simulation output was overlaid onto minimum-bias \ppb data to get events with generated jets and correct underlying events. Several samples of events were produced for different intervals of generator-level $R=0.4$ jet $\pT$, besides the MB samples, in order to obtain good statistics precision over large range in charged particle $\pT$. The final \pp (\ppb) results were obtained by merging all \pp (\ppb) samples, while each sample was scaled by the cross section of the jet production in the appropriate jet $\pT$ range.

Charge particle tracks are reconstructed in the ATLAS Inner Detector \cite{Aad:2010ac}. Tracks are measured using a combination of silicon pixel detector (Pixel), silicon microstrip detector (SCT), and a straw tube transition radiation tracker (TRT), all immersed in a 2\,T axial magnetic field. Charged particles typically traverse 3 layers of silicon pixel detectors, 4 layers of double sided microstrip detector, and 36 straws. 

The minimum $\pT$ of tracks used in this analysis is 4\,\GeV. Each track is required to have at least 1 hit in the Pixel detector, a hit in the innermost layer if such hit is expected by the tracking model, and at least 6 hits in the SCT. Additionally, tracks with $\pT>10$\,\GeV are required to produce at least 8 hits in the TRT. These requirements select tracks with good $\pT$ resolution and suppress the contribution of poorly reconstructed tracks. Yet they limit the analysis to the coverage of the TRT detector, thus all tracks are selected with $|\eta|<2$. 

To ensure that the tracks originate from the event vertex, the transverse impact parameter, $\dz$, and $\zz$ ($z_0$ is the longitudinal impact parameter) are required to be less than 1.5\,mm. Tracks selected for the analysis are also required to satisfy the conditions on the impact parameters significances: $|\dz/\sigma_{\dz}|<3$ and $|\zz/\sigma_{\zz}|<3$. The $\dz$ and $\zz$ parameters and their uncertainties are estimated by a vertex finding algorithm.

To reduce the amount of fake tracks at high $\pT$, all tracks with $\pT>15$\,\GeV are required to match anti-k${}_\mathrm{T}$ jets with the distance parameter $R=0.4$. The main interest of this analysis are spectra of charged hadrons. Thus, electrons and muons coming from electroweak decays of heavy bosons are subtracted from the measured spectra.

The raw charged particle spectra in \ppb and \pp collisions are corrected for fake tracks and secondary particles, for limited momentum resolution, and for tracking inefficiency. These corrections are functions of $\pT$ and rapidity in center-of-mass frame ($\ystar$) and are estimated using corresponding MC samples. First, the raw spectra are corrected for the fraction of tracks associated to secondary particles and for the fraction of tracks which cannot be associated to any particles in the MC, i.e. fake tracks. The fraction of secondary particles is at most 1\% at $\pT=4$\,\GeV. The fraction of fake tracks strongly depends on $\pT$ and at the highest measured $\pT$ reaches 13\%. It also depends on rapidity. Next, the resulting spectra are unfolded using the iterative Bayesian unfolding \cite{BayesUnf} to correct for the finite detector $\pT$ resolution. The first iteration is sufficient, further iterations do not change the results by more than 2\% even at the highest $\pT$. The spectra are then corrected for particle loss in the reconstruction by the track reconstruction efficiency.

Once the differential \pp cross section at $\sqs=2.76$ and 7\,\TeV have been measured, the differential \pp cross section at $\sqs=5.02$\,\TeV is obtained by interpolation. The interpolation is proportional to $\ln(\hspace*{-0.08cm}\sqs)$ and it is performed for every $\pT$ bin in each rapidity interval used in the \ppb analysis.

Systematic uncertainties are evaluated by varying individual sources within the ranges of their uncertainties and comparing the results to the results of the default analysis. Tracking and vertex pointing cuts contribute up to 5\%. The uncertainties on the correction for fake tracks and on the correction for track reconstruction efficiency contribute up to 10\% and 15\% at the highest $\pT$, respectively. Uncertainty on the momentum resolution of the Inner Detector tracking system adds a significant contribution of 12\% to the systematic uncertainties at high $\pT$. Difference up to 5\% was found between parts of the run with opposite directions of the \textit{p} and Pb beams in the LHC due to different jet trigger performance. The calculation of $\ystar$ is made with the assumption that all particles are pions, this introduces an uncertainty of 1\%. The uncertainty related to the description of the inactive detector material is found to vary between 1--7\% depending on $\eta$. 

The uncertainty on the event selection is accounted for in the uncertainty of the corresponding $\avgTpb$ values. The luminosities of the \pp data samples have an uncertainty of 3.2\% and 4\% for the measurements at $\sqs=2.76$ and 7\,\TeV respectively. Possible distortion introduced by the interpolation algorithm reaches up to 12\%. It is estimated as the difference between $\sqs$ and $\ln(\hspace*{-0.08cm}\sqs)$ interpolations. For $\rppb$, most of the systematic uncertainties between \ppb and \pp contribute to the final ratio, since the $\ystar$ variable corresponds to different areas in the detector, as the two systems are in two different centre-of-mass frames. 

\section{Results}

\begin{figure}[b!]
	\centering
	\includegraphics[height=0.3\textheight]{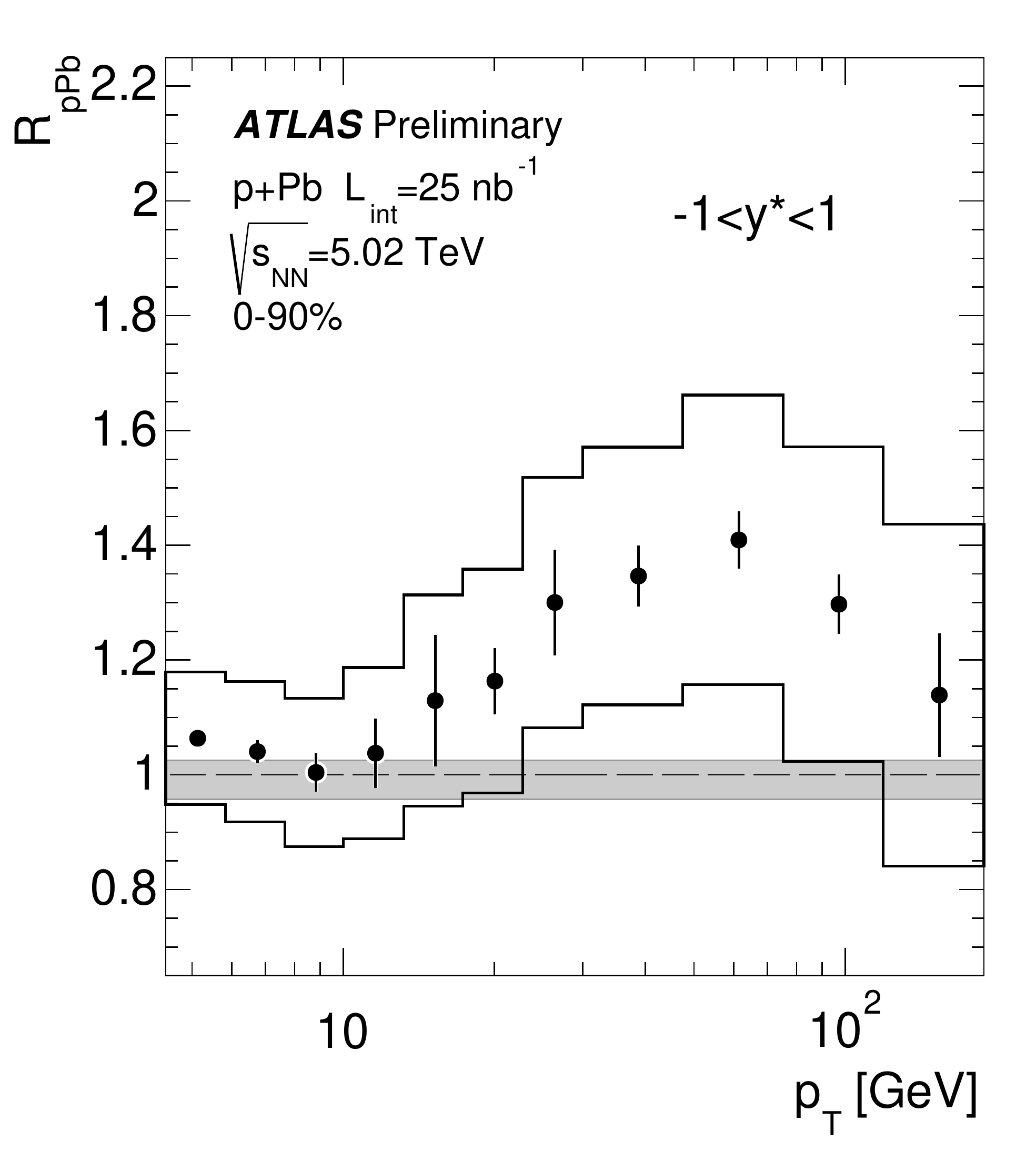} \includegraphics[height=0.3\textheight]{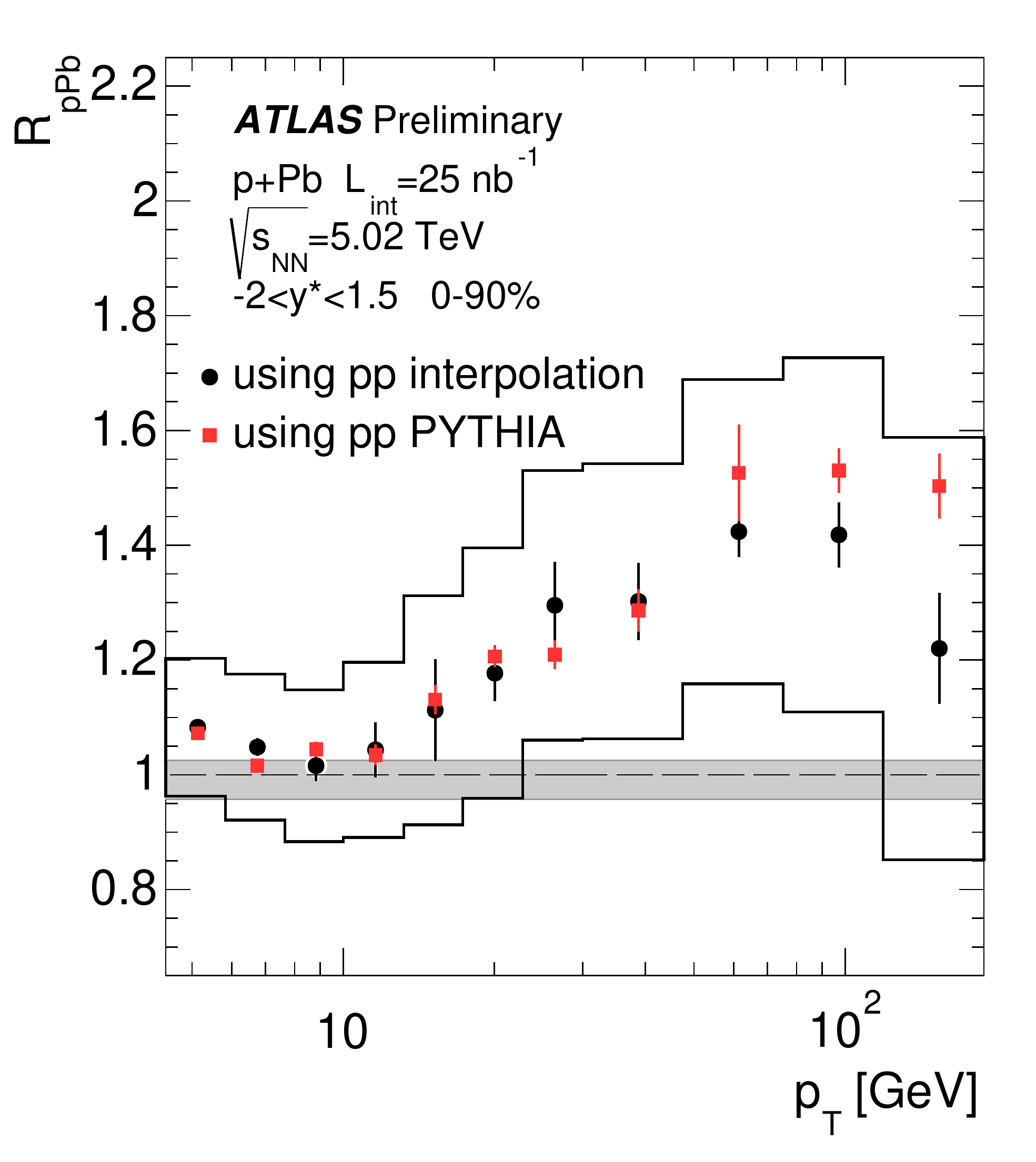}
	\caption{The $\rppb$ distributions as a function of $\pT$ in the 0--90\% centrality class. Left: $\rppb$ measured for the $|\ystar|<1$. Right: $\rppb$ with interpolated \pp reference (black points) and with simulated \pp reference (red squares). Vertical bars indicate the statistical uncertainty of the measurement. Systematic uncertainties are plotted with step lines encompassing each point, the systematic uncertainties of $\avgTpb$ are not included in the bands and shown with the gray-shaded bands at unity.}
	\label{rppb_mb_pyth}
\end{figure}

\begin{figure}[t!]
	\centering
	\includegraphics[height=0.3\textheight]{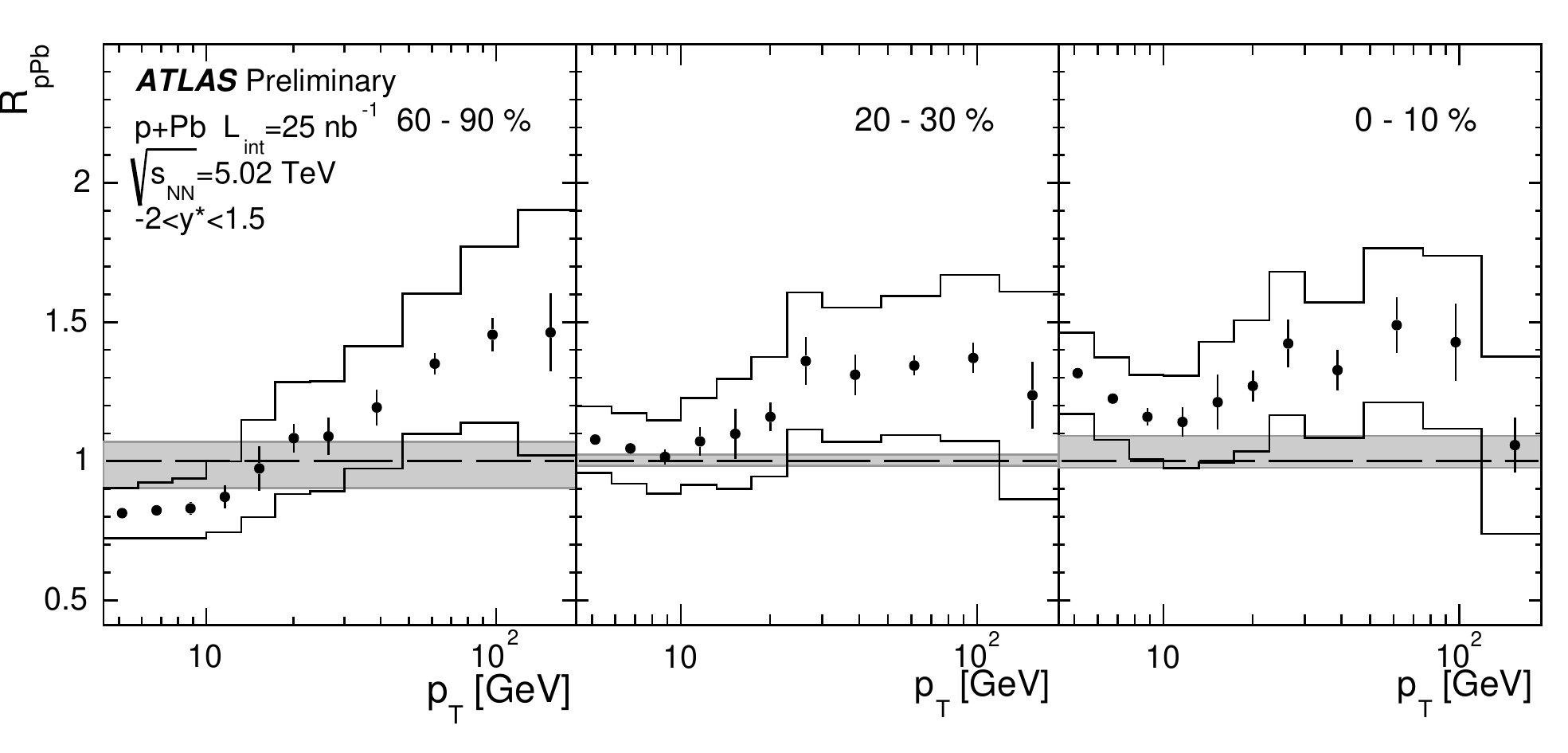}
	\caption{The $\rppb$ distributions as a function of $\pT$ measured for the 60--90\% (left), 20--30\% (middle) and 0--10\% (right) centrality class. Vertical bars indicate the statistical uncertainty of the measurement. Systematic uncertainties are plotted with step lines encompassing each point, the systematic uncertainties of $\avgTpb$ are not included in the bands and shown with the gray-shaded bands at unity.}
	\label{rppb_3c}
\end{figure}

The final charged particle spectra are MB spectra (including scaled jet triggered spectra). The \ppb yield is normalized per event, while the \pp yields are normalized by corresponding luminosity and interpolated to the same center-of-mass energy as \ppb sample.

Left panel of Fig.~\ref{rppb_mb_pyth} shows $\rppb$ in the rapidity interval $-1<\ystar<1$ for centrality class 0--90\%. The figure shows unexpected increase of $\rppb$ with increasing $\pT$, compared to the lower $\pT$ values. The $\rppb$ distribution reaches the maximum of approximately 1.4. This result is comparable with the CMS measurement \cite{CMS-PAS-HIN-12-017} and both measurements show the same trend at high $\pT$. 

Right panel of Fig.~\ref{rppb_mb_pyth} shows comparison of $\rppb$ with the interpolated \pp reference and with a reference obtained by pure simulation at $\sqs=5.02$\,\TeV. The increase of $\rppb$ at high $\pT$ is similar for these two distributions.

Figure \ref{rppb_3c} shows the $\rppb$ distributions for 3 centrality classes (0--10\%, 20--30\% and 60--90\%) and for the rapidity interval \mbox{$-2<\ystar<1.5$}. Especially in the 60--90\% centrality class, there is a clear trend of increasing enhancement towards higher $\pT$. In the more central classes the enhancement is presented as well.

\section*{Acknowledgements}
This research is supported by Charles University in Prague (projects UNCE 204020/2012, GAUK 500912 and GAUK 713612) and by INGO LG13009.

\bibliographystyle{elsarticle-num}
\bibliography{QM14-PetrBalek}

\end{document}